\documentclass[11pt,a4paper]{article}
\pdfoutput=1  

\usepackage[utf8]{inputenc}
\IfFileExists{lmodern.sty}{\usepackage[T1]{fontenc}\usepackage{lmodern}}{}
\usepackage{microtype}

\usepackage[a4paper,top=2.9cm,bottom=3.1cm,left=3.2cm,right=3.2cm]{geometry}
\usepackage{amsmath,amssymb,amsthm}
\usepackage{acronym}
\usepackage[dvipsnames]{xcolor}
\usepackage{fancyhdr}
\usepackage[numbers,sort&compress]{natbib}
\usepackage{orcidlink}
\usepackage{hyperref}

\definecolor{linkblue}{HTML}{18497E}
\definecolor{rulegray}{HTML}{8C8C8C}
\hypersetup{
  colorlinks = true,
  linkcolor  = linkblue,
  citecolor  = linkblue,
  urlcolor   = linkblue,
  pdftitle   = {Data Citation for Large Language Models: A Challenge},
  pdfauthor  = {Gianmaria Silvello},
  pdfsubject = {Data citation, provenance and credit attribution in large language models},
  pdfkeywords= {data citation, credit distribution, provenance, large language models,
                retrieval-augmented generation, knowledge graphs, data quality}
}
\makeatletter
\g@addto@macro{\UrlBreaks}{\UrlOrds}
\makeatother
\newcommand{\preprintrunner}{Preprint of the article to appear in the ACM Journal of Data and Information Quality}

\fancypagestyle{plain}{%
  \fancyhf{}%
  \fancyfoot[L]{\footnotesize\itshape\color{rulegray}\preprintrunner}%
  \fancyfoot[R]{\footnotesize\thepage}%
}

\theoremstyle{definition}

\theoremstyle{plain}

\newenvironment{researchquestion}
  {\par\addvspace{1.05\baselineskip}%
   \noindent\begin{minipage}{\linewidth}%
   {\color{rulegray}\hrule height 0.7pt}%
   \vspace{0.6\baselineskip}%
   \normalsize\bfseries\noindent\ignorespaces}
  {\par\vspace{0.6\baselineskip}%
   {\color{rulegray}\hrule height 0.7pt}%
   \end{minipage}\par\addvspace{1.05\baselineskip}}

\acrodef{KG}[KG]{Knowledge Graph}

\begin{document}

\title{Data Citation for Large Language Models:\\[2pt] A Challenge%
\thanks{Preprint of an article to appear in the \emph{ACM Journal of Data and
Information Quality} (JDIQ). The version of record will be available at
\url{https://doi.org/10.1145/3838808}. Released under a CC BY 4.0 license.}}

\author{%
  Gianmaria Silvello\,\orcidlink{0000-0003-4970-4554}\\[3pt]
  \normalsize Department of Information Engineering, University of Padua\\
  \normalsize Via G. Gradenigo 6/b, 35131 Padua, Italy\\
  \normalsize \texttt{gianmaria.silvello@unipd.it}
}
\date{}

\maketitle
\thispagestyle{plain}

\begin{abstract}
\noindent
Large language models increasingly mediate access to information, and a growing
body of work asks whether they cite the sources behind their outputs. That work
treats citation as a verification device and applies it to textual documents.
Scholarly citation serves two further functions, credit and provenance, and it
applies to data as much as to text. This paper argues that data citation for
large language models is an open challenge, distinct from document-level
citation grounding and harder to solve. We ask how such models should cite data
so that outputs stay verifiable, provenance stays traceable, and credit reaches
data creators and curators. We set out three research directions. Training data
attribution has to turn influence estimates into references for corpora
absorbed into model parameters. Data citation at inference time has to identify
datasets, subsets, and query results at the right granularity and fixity.
Citing knowledge graph facts has to define what a reference to a single triple
denotes and how credit propagates along provenance. Progress on all three
depends on joint work across the database, information retrieval, knowledge
representation, and artificial intelligence communities.

\end{abstract}

\section{Motivation}
Large Language Models (LLMs) are rapidly becoming primary interfaces for information access, synthesis, and decision support. An LLM draws on many sources: the training corpora absorbed into its parameters and, at inference time, retrieved documents (as in Retrieval-Augmented Generation (RAG)), \acp{KG} used for grounding, and tool-augmented or agentic workflows (database queries, search and KG-traversal APIs, code execution). The origin of an answer is therefore often a dynamic process spanning several sources rather than a single retrieved chunk. A growing body of work asks whether LLMs reliably cite the sources behind their outputs \cite{WallatHRA25, dassen2026, ding2025}; this line of research treats \textit{citation} as a verifiability mechanism: 
\emph{Does the model point to a source that supports its claim?}

This framing reflects only one aspect of a deeper issue. In scholarly work, citation has at least three functions: \textit{verification} (allowing readers to check claims), \textit{credit} (acknowledging intellectual contributions), and \textit{provenance} (tracing the origin of information) \cite{Liu1993, Borgman12}. Current LLM citation efforts focus almost entirely on verification for textual documents, largely neglecting \textit{data}: training datasets, structured contextual data, and the \acp{KG} increasingly used to ground LLM reasoning. This paper poses the following research question:

\begin{researchquestion}
How should LLMs cite data so as to ensure verifiability,
provenance tracking, and fair credit attribution to data creators and curators?
\end{researchquestion}

This question poses a challenge that is both technically harder and, we believe, socially more consequential than document-level citation grounding; this is particularly true because solving the ``data problem'' carries over to text in an almost straightforward way.  Data citation in traditional scholarly publishing is still an unsolved problem \cite{silvello2018, BunemanEtAl2020}; transposing it to the LLM setting amplifies every known difficulty and introduces new ones. The challenge sits at the intersection of database theory, information retrieval, knowledge representation, and AI ethics, and its investigation requires contributions from all these communities.

\section{Background}
The FORCE11 Joint Declaration of Data Citation Principles \cite{force11_2014} established data as first-class scholarly objects whose citations are human-understandable, machine-actionable, and support credit, provenance, and verifiability. Despite infrastructure progress -- DataCite \cite{datacite2025}, Scholix \cite{CousijnEtAl2018}, and the RDA recommendations for referencing dynamic data \cite{rauber_2025} -- adoption remains far from universal, and core problems persist: citing subsets and query results over evolving databases, ensuring fixity under updates, and generating complete, correct references at the right granularity \cite{BunemanEtAl2016, buneman2006}. Formal view-based models exist \cite{WuADS18, DavidsonBDMS17} but they are not pervasive, and even well-formed dataset references lack a common standard, so informal citations remain widespread \cite{lafia2023, federer2020, Irreraetal2023}. From an information-science standpoint, the challenge connects to the infrastructure layer that operationalizes credit and reuse: the FAIR principles, which require data to be findable and machine-actionable~\cite{wilkinson2016fair}; the CRediT contributor-role taxonomy, whose roles include data curation~\cite{niso2022credit}; and Make Data Count, which builds usage metrics intended to reward data creators~\cite{kratz2015making}. These frameworks define who deserves credit, how reuse is tracked and focus on metadata structure and tracking, yet none currently extends to data consumed inside LLM outputs.

Citing data from a curated database raises the question of how to distribute credit to the elements -- and ultimately the curators -- that produced it. \citet{dosso2020} formalized \textit{data credit distribution} on top of data citations and propagated it through a relational database's structure, later extended to lineage-based propagation along provenance chains \cite{dosso2022}. The concept of \textit{indirect citations} \cite{Fragkiadaki2014} and \textit{transitive credit} \cite{katz2014} generalizes this: if A contributes to B, the credit map of A feeds into B's. These mechanisms operate only within curated, provenance-aware databases.

\citet{huang-chang-2024-citation} argue that a comprehensive citation mechanism for LLMs must cover both non-parametric (retrieved) and parametric (training-internalized) content, a distinction that maps onto the separation of training-data attribution from RAG and KG citation developed in this paper. Recent work on RAG attribution shows that LLMs frequently misattribute sources: a cited document may not support the claim or may not be the one the model relied on \cite{WallatHRA25}, citation hallucination can be detected from a model's internal computations \cite{dassen2026}, and users over-trust answers that carry citations even when those citations are random \cite{ding2025}. This line of work targets factual grounding and citation \emph{faithfulness} -- whether a generated claim is supported by some source -- over textual documents. Data citation asks a different question: \emph{which} data objects should be cited, and how credit should flow to their creators and curators. Critically, all of this work addresses document citations; to our knowledge, data citations remain largely absent from the evaluation landscape.

Finally, LLMs and KGs are increasingly integrated \cite{pan2024}: KGs ground LLMs to reduce hallucination, especially in neurosymbolic approaches where symbolic knowledge constrains and explains neural outputs. Yet how an LLM should \textit{cite} the KG facts it uses, and how credit should reach their contributors, remains unaddressed.

\section{Challenges and Research Directions}

We identify three interconnected challenges, each of which opens concrete research directions.

\paragraph{Training Data Attribution.} Datasets used to train LLMs are absorbed into billions of parameters and are, in practice, lost to citation. The training data attribution problem asks: given a model output, which training data points contributed to it? Existing approaches include influence functions \cite{pmlr-v70-koh17a} and data Shapley values \cite{pmlr-v97-ghorbani19c}; large-scale audits separately document how licensing and attribution metadata are lost as data are absorbed into training pipelines~\cite{Longpre2024}. However, these methods are computationally expensive, approximate, and do not produce anything resembling a \textit{citation} in the scholarly sense.

A training-data citation would map an output to the training items that shaped it, each with a contribution weight; defining it rigorously and efficiently for models with billions of parameters and trillions of tokens is open. The problem becomes even more challenging when the training corpus includes structured datasets in addition to unstructured text, as it raises the question of how to appropriately reference a specific table, column, or tuple. We further emphasize that addressing this challenge is realistically feasible only for open models with documented training corpora.

\paragraph{Data Citation in LLMs.}
At inference time an LLM acquires data through several mechanisms, each raising the same citation problem, open even in traditional settings. RAG is the prototypical case. When structured datasets are provided as grounding context in a RAG pipeline, writing a proper reference for a dataset requires specifying the creator, title, version, persistent identifier, and access information -- metadata that is often incomplete or absent \cite{DelgadoQuiros2024}. If a model uses a specific subset of a dataset (e.g., rows matching a query predicate), the citation should reflect that subset with appropriate granularity, yet, to our knowledge, current LLM-based systems do not support this. Beyond retrieval, the citable object is frequently not a static document but a query and its result over a possibly versioned source, which makes granularity and fixity harder to pin down.
Furthermore, the scarcity of well-formed data citations in the textual corpora on which LLMs are trained means that models have seen far fewer examples of proper data references than document references. This training distribution imbalance biases LLMs toward informal or absent data attribution, in what we conjecture is a self-reinforcing deficit.

\paragraph{Citing Knowledge Graph Facts.} When an LLM uses a fact -- e.g., \textit{(Aspirin, treats, Headache)} -- three questions arise. (i) \textbf{What is a citation to a fact?} A triple has no bibliographic metadata, and citing the KG as a whole (``according to Wikidata'') conflates millions of independent assertions. (ii) \textbf{Provenance is essential:} a fact may be extracted from a paper, curated by an expert, inferred, or aggregated, and this determines who is credited; provenance-aware representations such as nanopublications \cite{FabrisKS19} and PROV-O-annotated named graphs \cite{dibowski2024} exist but are not integrated into LLM pipelines. (iii) \textbf{Credit must be distributed:} when a fact derives from several sources, the credit for using it should be split among them in proportion to their contribution, and when facts share upstream sources this becomes a network-propagation problem over the provenance graph. This is the \textit{credit distribution} problem \cite{dosso2020, dosso2022} (with transitive credit \cite{Fragkiadaki2014, katz2014}), now at the scale, heterogeneity, and dynamism of KGs in LLM pipelines, which existing solutions do not address.

A comprehensive solution must co-design five elements: a unified citation model spanning documents, datasets, and KG facts at different granularities (whole dataset, subset, individual triple); provenance-aware LLM architectures that preserve and expose the link between generated tokens and their data sources, whether these enter through training, retrieval, tool calling or symbolic grounding; credit-distribution mechanisms operating at scale over heterogeneous provenance graphs to distribute credit fairly among data creators, curators, and contributors; evaluation benchmarks for data citation in LLM outputs, analogous to document-citation correctness and faithfulness benchmarks but for (semi-)structured data; and standards and infrastructure extending existing frameworks (FORCE11, DataCite, Scholix, RDA) to the LLM context to make data citations in model outputs machine-actionable. These stages are interdependent rather than merely sequential. In the RAG case, a citation can be emitted only if the architecture has retained the link from the generated span back to the retrieved tuple or query result (provenance); credit can be distributed only if that citation identifies the data object at the right granularity (citation model); and neither can be assessed without a benchmark that scores data-level rather than document-level citations (evaluation). A gap at any one stage degrades the others.

Among these challenges, data citation in LLM systems is the first to tackle and the most immediately tractable: the data sources are available at inference time, and citations can be anchored to retrievable, versioned artifacts or to the queries that produced them, with RAG the natural entry point. It is also arguably the most pressing, because as LLM systems become the standard way to ground outputs, the lack of proper data citation risks normalizing large-scale consumption of structured data without attribution -- a gap that will only grow harder to close.

Data citation must therefore work not only in traditional publishing but in the AI systems that increasingly mediate access to knowledge. This requires the DB, IR, KR, and AI communities to jointly develop the models, algorithms, benchmarks, and standards above, so that knowledge integrity remains traceable at inference time.

\section*{Acknowledgments}
The work was supported by the HEREDITARY project, as part of the EU Horizon
Europe program under Grant Agreement 101137074.

\end{document}